\documentclass[twocolumn,showpacs,preprintnumbers,amsmath,amssymb]{revtex4}

\usepackage{graphicx}
\usepackage{dcolumn}
\usepackage{bm}

\begin{document}

\preprint{APS/123-QED}

\title{Temperature dependence of trapped magnetic field in MgB$_2$ bulk superconductor
}

\author{R. V. Viznichenko}
\email[\\Electronic mail: ]{vrv@imp.kiev.ua}
\author{A. A. Kordyuk}
\affiliation{Institute of Metal Physics, 03142 Kyiv, Ukraine}

\author{G. Fuchs, K. Nenkov}
\author{K.-H. M\"{u}ller}
\affiliation{Institute for Solid State and Materials Research, Dresden, Germany}

\author{T. A. Prikhna}
\affiliation{Institute for Superhard Materials, Kyiv, Ukraine}

\author{W. Gawalek}
\affiliation{Institut f\"{u}r Physikalische Hochtechnologie, Jena, Germany}

\date{\today}

\begin{abstract}
Based on DC magnetization measurements, the temperature
dependencies of the trapped magnetic field have been calculated for
two MgB$_2$ samples prepared by two different techniques: the
high-pressure sintering and the hot pressing. Experimentally measured
trapped field values for the first sample coincide remarkably well
with calculated ones in the whole temperature range. This proves, from
one side, the validity of the introduced calculation approach, and
demonstrates, from another side, the great prospects of the hot
pressing technology for large scale superconducting applications of
the MgB$_2$.
\end{abstract}

\pacs{74.25.Sv, 74.70.Ad}  
\maketitle


Magnesium diboride is a new promising superconducting material with a critical temperature of about 40 K.
High $T_c$ values and simple chemical composition of MgB$_2$ have made it an interesting object 
for applied investigations with great perspectives of use  in superconducting 
motors, flywheels and bearings. The key parameter for such applications is the maximum trapped field 
in the sample
and its temperature dependence. This, in turn, is closely connected both with the critical current density 
and with the size of superconductor.

Since the very discovery of superconductivity in MgB$_2$ the
researchers dealt with tiny samples which were not suitable for large
scale applications. Now, the situation is changed and several
techniques have been elaborated to produce high quality MgB$_2$ bulk
polycrystalline samples of a few centimeters in diameter and with the
critical current densities up to $10^5$ - $10^6$ A/cm$^2$ \cite{ref2, ref2a}.

It is remarkable that, in contrast to the high temperature
superconductors, the grain boundaries in bulk magnesium diboride
superconductors do not act as weak links \cite{ref3, ref3a}. This significantly
simplifies the growth of bulk samples suitable for large scale applications.

In this letter we focus on an important property of the bulk
superconductor -- the trapped magnetic field. We calculate its
temperature dependencies from DC magnetization data using an
iteration approach. We show that the trapped field values measured
experimentally are in a perfect agreement with the calculated ones.
This makes us able to predict the expected values of the trapped
magnetic field in a sample which is not yet grown to an appropriate
size but which preparation technique looks the most promising today.
\begin{figure}
\includegraphics[width=8.5cm]{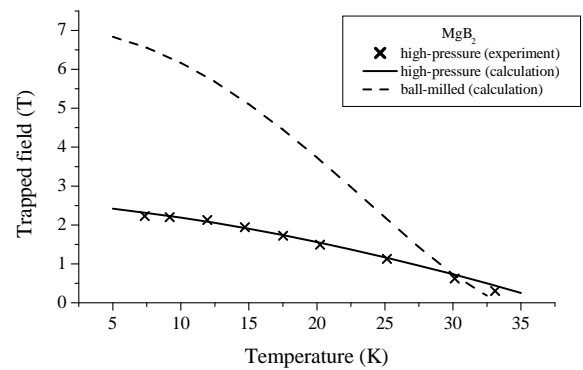}
\caption{\label{fig:1} Temperature dependencies of maximum trapped field
in bulk MgB$_2$  samples prepared under high-pressure sintering and ball-milled techniques.
}
\end{figure}

We have studied two MgB$_2$ bulk
samples prepared by different techniques. One of them was sintered under high pressure
 as described in \cite{ref4}. The sample has an uniform polycrystalline structure
 without cracks and has 28~mm in diameter and 11~mm in height.  
The trapped magnetic field in the centre of the sintered sample was measured 
by a Hall probe for different temperatures from 6 to 33 K. The resulting temperature
dependence of the trapped field has a negative curvature and 
trends to saturation at low temperatures (see Fig.~1, cross symbols).
The other sample was prepared by ball milling of Mg and B powders at ambient
 temperatures followed by hot pressing \cite{ref5}, which we call "ball-milled".  
It consists of spherical nanocrystalline grains 
about 40 - 100 nm in size, that distinctly improve pinning due to the large number of grain boundaries.
For this sample the trapped field can not be 
measured because of its small size. 

To calculate the trapped field one should know temperature and field dependencies 
of the critical current density $J_c(H, T)$ of the both superconductors. 
For DC magnetization measurements small bar shaped
pieces of the both samples have been used.
Then, the $J_c(H)$  curves were obtained from the magnetization loops
$M(H)$ using the conventional expression $j_c(H)=20\cdot \Delta M (b - b^2/3l)$, where $\Delta M$ is
the difference of the magnetization (in emu/cm$^2$) measured for ascending and descending applied field, $b$ and $l$
are the sample width and length in cm, respectively, and $j_c$ is obtained in A/cm$^2$.

For the high-pressure sintered sample a set of $J_c(H)$ curves presented in Fig. 2 
in logarithmical scale are straight lines  for low fields up to the level of $J_c~=~10^4$~A/cm$^2$.
This means an exponential decay of critical current density with increasing field
\begin{equation}
J_c(H, T) = J(T)exp(-H/b(T)) ,
\label{eq:1}
\end{equation}
where $J(T)$  and $b(T)$ are temperature dependent coefficients.
\begin{figure}
\includegraphics[width=8.5cm]{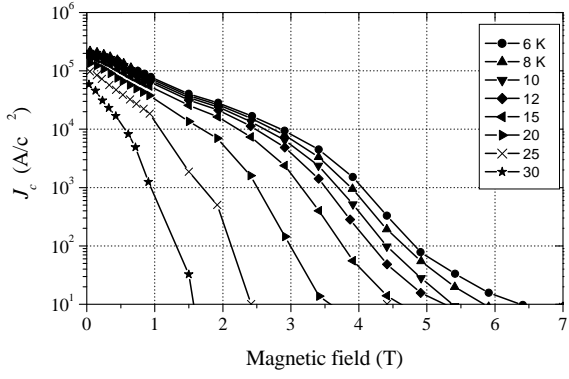}
\caption{\label{fig:2} Field dependencies of critical current density at different temperatures for
high-pressure sintered MgB$_2$  sample. $J_c(H)$ curves were obtained from DC magnetization data.}
\end{figure}

The dependencies of $J(T)$  and $b(T)$ for the sintered sample, shown in Fig. 3 by circles,
 can be well fitted to the expressions
\begin{equation}
b(T) = b_0[1-(T/T_c)^2] ,
J(T) = J_0[1-(T/T_c)] .
\label{eq:2}
\end{equation}
Here $T_c$ = 37.5 K is the critical temperature, whereas $b_0$~=~0.9 T and 
$J_0$ = $3.1\times10^5$ A/cm$^2$ are fitting parameters, which characterize an effective field
of supercurrent decay and a maximum critical current density at zero field and temperature, respectively.

For the ball-milled MgB$_2$ sample $J_c(H,T)$ can be described by Eq.~(\ref{eq:1}) too 
(see Fig.~4 in \cite{ref5}), but the functions $b(T)$ and $J(T)$ are different:%
\begin{equation}
b(T) = b_0[1-(T/T_c)^2]^{3/2},
J(T) = J_0[1-(T/T_c)^2]^{3/2},
\label{eq:3}
\end{equation}
where the corresponding parameters are 
$T_c$ = 34 K, $b_0$~=~2.78~T and $J_0$ = $8.5\times10^5$ A/cm$^2$ (see Fig.~3, squares).
 
The obtained results show that the critical current density strongly depends on temperature. 
For the high-pressure sintered sample the values of $J(T)$ and $b(T)$ are about 2.5 times lower than those of
for the ball-milled sample. From the other hand, the latter sample has a much stronger power dependence of
$J(T)$ and $b(T)$ on temperature.
\begin{figure}
\includegraphics[width=8.5cm]{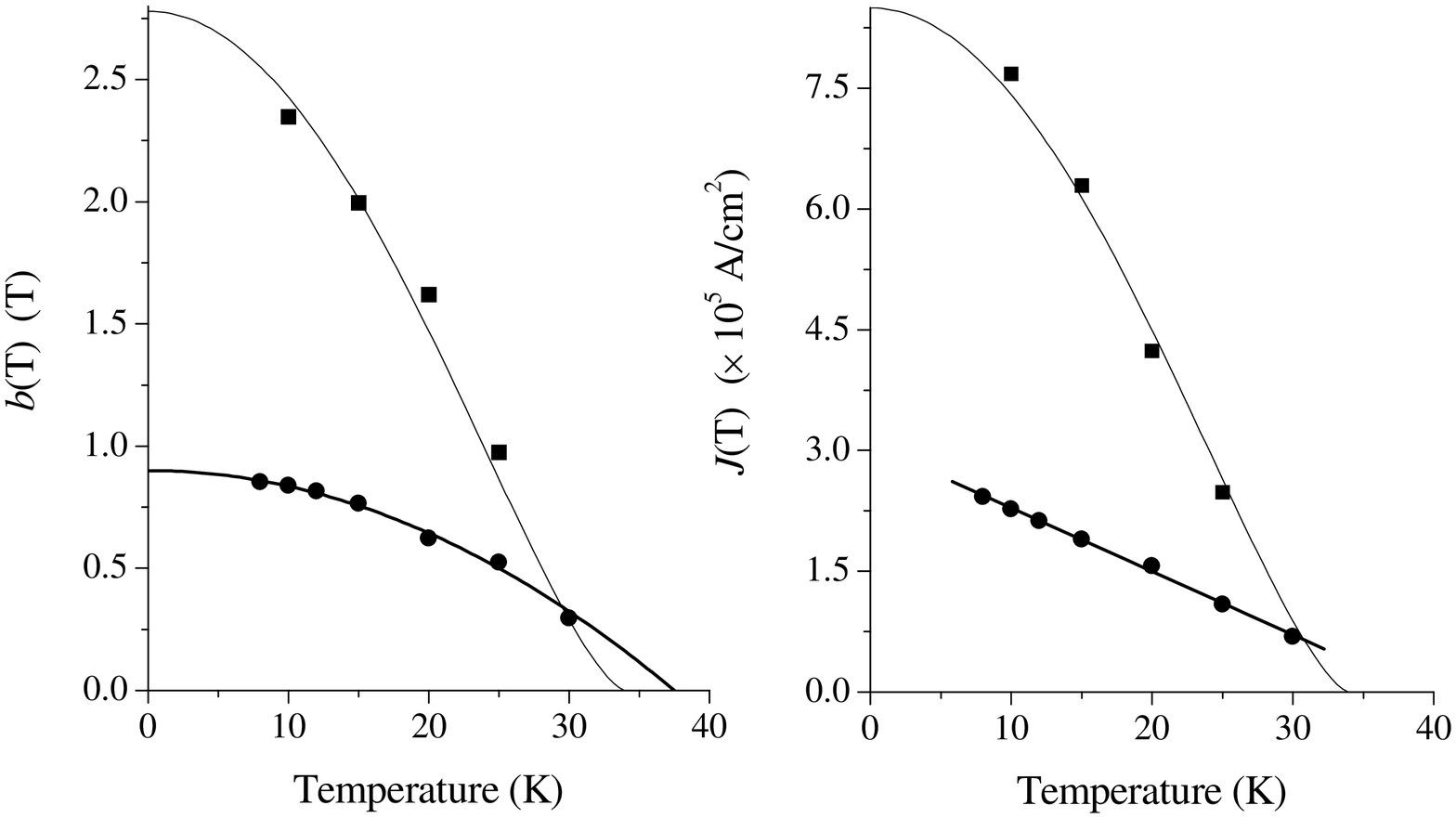}
\caption{\label{fig:3} $J(T)$ and $b(T)$ coefficients of $J_c(H)$ curves (see Eq.~\ref{eq:1}) for MgB$_2$ bulks. 
Symbols represent experimental data for high-pressure sintered (circles) and ball-milled (squares) samples. 
Solid lines represent fitting results, which is described by formulas (\ref{eq:2}) and (\ref{eq:3}) 
for sintered and ball-milled
 samples respectively.}
\end{figure}

The further calculation of the trapped field from a known $J_c(H, T)$
function is a nontrivial task. According to the Biot-Savart low, the
magnetic field $\bm{H}$ generated by the supercurrent $\bm{J_c}$
flowing in the volume of sample $V$ is
\begin{equation}
\bm{H(r)} = \frac{1}{c} \int_{V} \bm{J_c(H,\rho)} \times \frac{\bm{r - \rho}}{|\bm{r - \rho}|^3} d^3\rho.
\label{eq:4}
\end{equation}
As far as $\bm{J_c}$ is a function of $\bm{H}$, one needs to solve this integral equation in order to calculate the
trapped field values.

To simplify the problem we assume the homogeneity of the sample, i.e.
(1) that the critical current density does not depend on coordinate
$\rho$ explicitly and (2) that the currents flow along concentric
circles.
Then, the field component normal to the sample surface, $H_z$, 
can be written in cylindrical coordinates $\bm{r}=(r,\phi,z)$ and $\bm{\rho}~=~(\rho,\psi,\zeta)$ as follows
\begin{equation}
H_z(r) = \frac{1}{c} \int_{V} J_c(H_z) f(r, \rho, \psi, \zeta)\rho d^3\rho,
\label{eq:5}
\end{equation}
where $f(r,\rho,\psi,\zeta)=\frac{\rho - r\cos\psi}{[(r - \rho\cos\psi)^2 + (\rho\sin\psi)^2 + (z - \zeta)^2]^{3/2}}$, 
$d^3\rho = d\rho d\psi d\zeta$. The resulting field profile has a cylindrically symmetric 
form with the maximum at $r=0$. 

The integral equation~(\ref{eq:5}) can be solved for $H(r)$ numerically using an 
iterative procedure \cite{ref6}
\begin{equation}
H_{i+1}(r) =\frac{1}{c} \int_{V} J_c(H_i) f(r, \rho, \psi, \zeta)\rho d^3\rho,
\label{eq:6}
\end{equation}
until $|H_{i+1} - H_i|/H_i < \varepsilon$ . 
Here the obtained profile is used to calculate current in the next step of iterations. We chose
 an initial profile $H_0(r) = const$ to start the iterations.

One should note that we calculate the trapped field profile numerically in each step and 
can not define $H(r)$ function in every point. Therefore, we should discretize Eq.~(\ref{eq:5}) and divide 
the cylindrical 
sample into $N$ concentric tubes of width $\Delta = \frac{R_s}{N-1}$ and then calculate the field
 in each point $r_k, k = 0 \dots N-1$ as a sum
of fields from all tubes
\begin{equation}
H(r_k) = \frac{\Delta}{c} \sum^{N-1}_{j=0}\int J_c(H(\rho_j)) f(r_k, \rho_j, \psi, \zeta) \rho_j d\psi d\zeta,
\label{eq:7}
\end{equation}
where $r_k = k\Delta$, $R_s$ is the radius, $L$ is the height of the cylindrical sample and
the integration here is made over $\psi$ from $0$ to $2\pi$ and over $\zeta$ from $-L$ to $0$. 

From equations (\ref{eq:7}) one obtains the field profile for a given $J_c(H)$ function using iteration (\ref{eq:6}).
In the case of $J_c(H) = const$ we will obtain a conical Bean profile with the maximum in the center of the sample.
To reach the accuracy of $\varepsilon=0.1~\%$  it is sufficient to make just 16 iterations with $N = 14$ tubes. 
Moreover, the method of calculation has no fitting parameter and use only experimental $J_c(H, T)$ data. 
The calculated temperature dependence of maximum trapped field for high-pressure sintered and ball-milled 
MgB$_2$ samples are represented on Fig. 1 by the solid and dashed lines respectively. 

We obtained a very good coincidence with experimental points for the high-pressure sintered sample. 
Remarkably, the trapped field, which is determined by the current distribution in the whole sample volume 
can be correctly calculated on the base of $J_c$ measured locally in a small piece of the sample.
The ball-milled (nanocrystalline) sample was found to have much higher trapped fields than the sintered sample.
Thus, the development of large size nanocrystalline MgB$_2$ samples is very attractive for future applications.

To sum up, we calculated the temperature dependence of the trapped magnetic field for two MgB$_2$ bulk samples
prepared under different techniques. We used DC magnetization data and an
iteration approach to solve the Biot-Savart equation. Experimentally measured trapped field values for the 
high-pressure sintered sample coincide excellent with calculated ones in the whole temperature range. The correctness of
the calculation allowed us to predict the expected values of the trapped field in the ball-milled sample and revealed
the great prospects of this technology for large scale applications of superconducting MgB$_2$.

\bibliography{MgB2_c}

\end{document}